\documentclass[aps,prl,twocolumn,showpacs,superscriptaddress,groupedaddress]
{revtex4}

\usepackage{dcolumn}   
\usepackage{bm}        
\usepackage{amssymb}   

\usepackage{amssymb,amsfonts,bm}
\usepackage{graphics,graphicx}
\usepackage{colordvi,amsmath,epsfig,color}
\usepackage[activeacute,english]{babel}
\begin{document}

\title{Understanding the instability of a vibrated granular monolayer}

\author{P. Maynar}
\author{M. I. Garc\'ia de Soria}
\author{J. Javier Brey}

\affiliation{F\'{\i}sica Te\'{o}rica, Universidad de Sevilla,
Apartado de Correos 1065, E-41080, Sevilla, Spain}

\date{\today}

\begin{abstract}
We investigate the dynamics of an ensemble of inelastic hard spheres
confined between two horizontal 
plates separated a distance smaller than
twice the diameter of the particles, in such a way that the system is
quasi-two-dimensional. The bottom wall is vibrating and, therefore, it
injects energy into the system
in the vertical direction 
and a stationary state is reached. It is found that, 
if the size of the plates is small enough, 
the stationary state is homogeneous. Otherwise, a cluster of
particles is developed. 
The instability is understood by 
using some effective hydrodynamic equations in the horizontal plane. 
Moreover, the theoretical prediction for the size of the system above
which it is unstable agrees very well with Molecular Dynamics
simulation results without any fitting parameter. 
\end{abstract}

\maketitle

A granular system is an ensemble of macroscopic particles, named
grains, that interact via inelastic collisions.  When two grains
collide, due to its macroscopic character, part of the kinetic energy
associated to the center of mass of the particles 
is dissipated exciting other degrees of freedom. Hence, besides the
fact that granular matter is ubiquitous in Nature, these systems are very
interesting from a theoretical point of view because they are
intrinsically out of equilibrium \cite{g03}. 

Experimentally, it is relatively easy to
generate stationary states by just vibrating the walls of the
container of the system. In these states, the energy injected in the
system is compensated by the energy dissipated at collisions. The
price to be paid is that the steady state is highly
inhomogeneous, as can be understood from the generalized Navier-Stokes
equations describing the dynamics of the system \cite{bdks98}. An
exception to this is a granular system placed in a vertically vibrated
quasi-two-dimensional (Q2D) shallow box. In this case, the stationary state
can be homogeneous when projected in the horizontal plane and, when observed
from above, the system behaves as a two-dimensional fluid. Actually,
in the last years, many experiments have been performed under these
conditions \cite{ou98, peu02, mvprkeu05, ou05,
  cms12, gs18}. Experiments can be carried out with or without a top lid,
being gravity in the last case the responsible of the Q2D confinement. 

Typically, the grains are confined between two plates separated a
distance $H$ smaller than twice the diameter of the particles, so that
they can not jump on to each other and the system can be considered to
be Q2D. The bottom wall is vibrated sinusoidally
with frequency $w$ and amplitude $A$ that is always much smaller than
the height of the system, i.e. $A<<H$. Most of the experiments are
performed with smooth plates although in some cases a rough one is
used \cite{peu02}. The particles gain energy in the vertical direction
through collisions with the bottom wall and it is transferred to the
horizontal direction via collisions between particles, where energy is
also dissipated. Experiments show that, for a wide range of the
parameters, the system reaches a homogeneous stationary state. On the
other hand, it is also observed that the system becomes unstable by increasing
the density and/or decreasing  the dimensionless parameter
$\Gamma\equiv\frac{Aw^2}{g}$, $g$ being the gravitational
acceleration \cite{ou98, mvprkeu05, ou05, cms12}. 
In particular, there is a regime in which an
aggregate surrounded by a hotter gas is formed. In the last years,
there have been many efforts to understand this phenomenology. For
example, it is known that, in the context of hydrodynamics, the
instability is triggered by a negative 
compressibility in the associated horizontal equation of state
\cite{ccdhmrv08, ka11} and some
of the phases have been characterized \cite{ou05,
  cms12}. Nevertheless, there are still many points that are not
clear. In particular,  what are the essential ingredients to have this
kind of instability?, why is the
compressibility negative?,  can the equation of state in the horizontal
plane be derived from a microscopic point of view? In any case, what
is the mechanism that stabilizes/destabilizes the system? 
In order to tackle these questions some two
dimensional (D) models 
have been considered. As the system is Q2D, it is expected 
that the actual 3D dynamics could be substituted by an effective 2D
dynamics in which energy is injected by some kind of mechanism as, for
example in the the stochastic 
thermostat \cite{wm96} or in the so-called $\Delta$ model
\cite{brs13}. Nevertheless, both models have been extensively
studied finding that the homogeneous stationary state that is reached
in the long time limit is always stable \cite{gmt13, gcv13,
  bbgm16}. Very recently, a new 2D model has been formulated in which
the homogeneous stationary state is unstable \cite{rsg18}, 
but the proposed microscopic dynamic is defined in terms of some 
parameters that, in principle, must be fitted. 

\begin{figure}
\begin{center}
\includegraphics[angle=0,width=0.8\linewidth,clip]{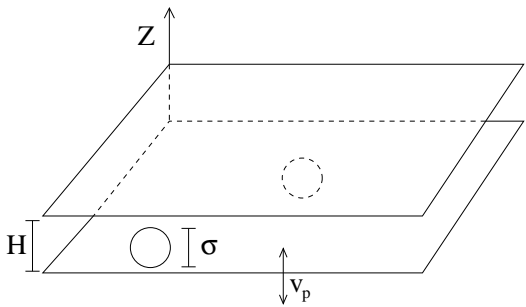}
\end{center}
\caption{Schematic representation of the model.}\label{monoFig}
\end{figure}

The objective in this paper is to introduce the simplest model that
captures the phenomenology of the experiments and that let us
understand the origin of the instability. 
To properly describe the energy transfer from the vertical to the
horizontal direction, we consider a simple 3D model: an ensemble of
$N$ inelastic hard spheres of mass $m$ and diameter $\sigma$, confined
between two flat planes located at $z=0$ and $z=H$. It will be assumed
that $H<2\sigma$ so that the system is Q2D (see Fig. \ref{monoFig}). 
The plates are square
shaped of area $L\times L$ with $L>>H$ and periodic boundary
conditions are used in the horizontal direction. The collisions
between the 
particles are inelastic, characterized by a constant (independent
of the relative velocity) coefficient of normal restitution,
$\alpha$ ($0<\alpha\leq 1$, being the collisions elastic for
$\alpha=1$). The collisions with 
the top wall are elastic, so that the horizontal component of the
velocity does not change and $v_z\to -v_z$. 
The bottom wall will
be modeled by a sawtooth wall of velocity $v_p$. This kind of wall
mimics the dynamics of a wall that moves sinusoidally 
in the limiting case $A\to 0$,
$w\to \infty$ with $Aw=v_p$. When a particle collides with the
bottom wall, its horizontal velocity remains unchanged while $v_z\to
2v_p-v_z$. This kind of
collision always injects energy into the system. The total
horizontal momentum of the system is, then, a constant of the motion. 

\begin{figure}
\begin{center}
\includegraphics[angle=0,width=0.55\linewidth,clip]{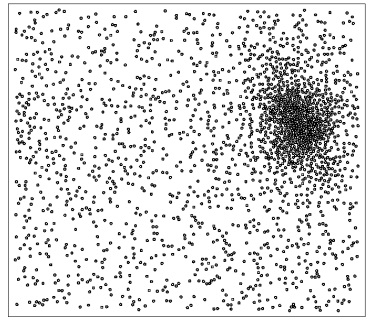}
\end{center}
\caption{Snapshot of a configuration of a MD simulation where the cluster is
  seen (the system is observed from above).}
\label{clusterFig}
\end{figure}

We have performed Molecular Dynamics (MD) simulations of the model
finding that it fulfills the desired conditions: for a wide class of
values of the parameters the system reaches a homogeneous stationary
state and, when increasing the density, a dense cluster 
surrounded by a gas is developed. In Fig. \ref{clusterFig} a snapshot
of a configuration of a MD simulation where the cluster is formed is shown (the system is
seen from above). The parameters of the simulations are $N=2000$,
$\alpha=0.9$, $H=1.5\sigma$, $L=115\sigma$ and
$v_p=0.01\left[\frac{2T(0)}{m}\right]^{1/2}$, where $T(0)$ is the
initial horizontal temperature. 

The objective now is to understand, first, the homogeneous
two-dimensional phase and,
second, the origin of the instability. To study the homogeneous phase,
we will admit, in the same spirit that in Ref. \cite{ka11}, the
existence of a closed
description in terms of the 2D density, $n$, the horizontal temperature,
$T$, and vertical temperature, $T_z$. The idea is that, as energy is
injected in the vertical direction and transfer to the horizontal
direction through collisions, this is the minimal number of variables
to understand the dynamics of the homogeneous 
system. Total momentum in the plane does not play any role in this
context as it is a constant of the motion. In 
the low-density limit, assuming that the system is very
thin, i.e. $\epsilon\equiv\frac{H-\sigma}{\sigma}<<1$, and that the
collisions between particles are nearly elastic, $1-\alpha<<1$, 
the evolution equations are 
\begin{eqnarray}
\frac{dT}{dt}&=&2\sqrt{\frac{\pi}{m}}n\sigma T ^{\frac{1}{2}}\left[-(1-\alpha)T
+\frac{\epsilon^2}{3}(T_z-T)\right], \label{ecT}\\
\frac{dT_z}{dt}&=&-\frac{4}{3}\sqrt{\frac{\pi}{m}}\epsilon^2n\sigma
                   T^{\frac{1}{2}} (T_z-T)+\frac{2v_p}{\epsilon\sigma}T_z. \label{ecTz}
\end{eqnarray}
Let us discuss briefly each term in the equations. The first term in
the right hand side of 
Eq. (\ref{ecT}) describes the energy loss due to inelastic collisions
between particles and the 2D expression has been taken
\cite{g03}. The second term is the kinetic energy transfer from the vertical to
the horizontal direction that is proportional to the difference of
temperatures times the thermal horizontal velocity. We have taken the
elastic limit calculated in \cite{bgm17} 
from kinetic theory tools. This term also appears in
Eq. (\ref{ecTz}) but with a different sign and multiplied by $2$ due to
the difference in the definition of temperature in terms of
energy. Finally, the second term in the right hand side of
(\ref{ecTz}) describes the energy injection due to the vibrating
wall. In this case, the term can be 
evaluated exactly as it is proportional to $v_p$ times the
pressure of the granular gas just above the vibrating wall
in the direction perpendicular to it \cite{br09}. In fact, these equations can be derived from a kinetic
equation for the monolayer by an expansion in $\epsilon$ and assuming
that the velocity distribution function is a Gaussian with two temperatures
\cite{bgm18}. The only difference is that the energy transfer terms
have a non-trivial dependence on $\alpha$ neglected in the present
case.

Eqs. (\ref{ecT}) and (\ref{ecTz}) admit only one stationary solution, $T_s$
and $T_{z,s}$, that can be easily calculated as functions of the
inelasticity, density and dimensionless heigh, $\epsilon$. From
Eq. (\ref{ecT}) the quotient of the stationary temperatures, 
$\gamma\equiv\frac{T_{z,s}}{T_s}$, is 
\begin{equation}\label{qEq}
\gamma=1+\frac{3(1-\alpha)}{\epsilon^2},  
\end{equation}
that is density independent. Moreover, it is always greater
than one, with equipartition holding only in the elastic limit. The stationary
horizontal temperature is
\begin{equation}\label{ts}
T_s=\frac{m}{4\pi}\left[\frac{\gamma v_p}{(1-\alpha)\epsilon
    n\sigma^2}\right]^2. 
\end{equation}
Remarkably, it is proportional to $n^{-2}$, so that the pressure in
the stationary state goes
as $n^{-1}$, and the compressibility of the non-equilibrium steady
state is negative. Note that this 
dependence is a direct consequence of the particular way in which
energy is injected and transferred in the monolayer.

In order to see that Eqs. (\ref{ecT}) and (\ref{ecTz}) describe 
correctly the dynamics of the system, we have performed MD simulations
measuring the time evolution of the horizontal and vertical
temperatures. We have considered different densities (below
$n\sigma^2=0.03$), 
heights (below $\epsilon=0.5$) and
coefficients of normal 
restitution (above $\alpha=0.9$), 
checking that the
system stayed always homogeneous. Of course, those cases in which the
instability showed up were
discarded. For all the considered initial
conditions (that was taken to be a Gaussian with two temperatures),
the numerical solution of the equations agrees very well with the
simulation results for all times and, in particular, with their
stationary values given by Eqs. (\ref{qEq}) and (\ref{ts}). This is
remarkable since there are 
not any adjustable parameter. 

A stability analysis of Eqs. (\ref{ecT}) and (\ref{ecTz}) shows that
the stationary solution is linearly stable. In fact, the matrix
associated to the dynamics of $\delta T\equiv T-T_s$ and 
$\delta T_z\equiv T_z-T_{z,s}$ has two eigenvalues, one of which is
always the slowest one and, in addition, goes to zero in the elastic
limit.  Then, there is a time scale in which
the dynamics is dominated by this slowest mode, so that $T_z$ can be considered
to be slaved to $T$ and a description in terms of only one
temperature, $T$, is valid. In the following, we will call it
``homogeneous hydrodynamic'' time scale. In this time scale, the evolution equation
for the temperature is 
\begin{equation}\label{deltaTEq}
\frac{d}{ds}\delta T\simeq -\beta\delta T, 
\end{equation}
where we have introduced the dimensionless time 
$ds=n\sigma\sqrt{\frac{2T_s}{m}}dt$ and $\beta$ is the absolute value of
the slowest eigenvalue, that is a known function
of $\alpha$ and $\epsilon$. 

Now, let us suppose that there are gradients in the horizontal
plane. The previous results make plausible a description in terms
of the 2D density, $n(\mathbf{r},t)$, 2D flow velocity,
$\mathbf{u}(\mathbf{r},t)$, and horizontal temperature, $T(\mathbf{r},t)$. The
evolution equations for these fields are assumed to be of the form 
\begin{eqnarray}
\frac{\partial n}{\partial t}&=&-\nabla\cdot(n\mathbf{u}), \label{nEq}\\
\frac{\partial\mathbf{u}}{\partial
  t}&=&-\mathbf{u}\cdot\nabla\mathbf{u}
-\frac{1}{nm}\nabla\cdot\mathbf{P}, \label{uEq}\\
\frac{\partial T}{\partial t}&=&-G(n,T)-\mathbf{u}\cdot\nabla T
-\frac{1}{n}(\nabla\mathbf{u}:\mathbf{P}+\nabla\cdot\mathbf{q}), \label{TEq}
\end{eqnarray}
where $\mathbf{P}$ is the pressure tensor and $\mathbf{q}$ the heat
flux. In the low-density limit and to Navier-Stokes order, we assume 
$P_{ij}=nT\delta_{ij}-\eta\left(\frac{\partial u_j}{\partial x_i}
+\frac{\partial u_i}{\partial
  x_j}-\delta_{ij}\nabla\cdot\mathbf{u}\right)$, 
and $\mathbf{q}=-\kappa\nabla T-\mu\nabla n$, 
where $\eta$ is the shear viscosity, $\kappa$ the (thermal) heat conductivity
and $\mu$ an additional transport coefficient called diffusive heat
conductivity that couples the heat
flux with the density gradients and that is peculiar to inelastic
collisions \cite{bdks98}. The unknown term $G(n,T)$ describes the homogeneous evolution of
the temperature. All the gradient contributions coming from the
cooling/heating rate are neglected. 

The system of equations (\ref{nEq})-(\ref{TEq})
admits a homogeneous stationary solution, $n(\mathbf{r},t)=n_s$, 
$\mathbf{u}(\mathbf{r},t)=\mathbf{0}$ and $T(\mathbf{r},t)=T_s$. The
two constants 
$n_s$ and $T_s$ must be related by
Eq. (\ref{ts}) that has to be equivalent to $G[n_s,T_s(n_s)]=0$.
The objective now is to study if this state is linearly stable. To do
that, it is convenient to  introduce the dimensionless deviations of the fields
around the homogeneous stationary solution
$\rho\equiv\frac{n-n_s}{n_s}$,
$\mathbf{w}\equiv\sqrt{\frac{m}{2T_s}}\mathbf{u}$ and
$\theta\equiv\frac{T-T_s}{T_s}$, as functions of the dimensionless
time scale, $s$, and dimensionless  space variable 
$\mathbf{l}\equiv n_s\sigma\mathbf{r}$. Let us also introduce the Fourier
components of these functions through 
$y_\mathbf{k}\equiv\int
d\mathbf{l}e^{-i\mathbf{k}\cdot\mathbf{l}}y(\mathbf{l})$ and let us
decompose $\mathbf{w}_{\mathbf{k}}$ into its parallel, 
$w_{\mathbf{k},||}\equiv\mathbf{w}_{\mathbf{k}}\cdot\frac{\mathbf{k}}{k}$,
  and transversal 
$w_{\mathbf{k},\perp}\equiv\mathbf{w}_{\mathbf{k}}\cdot\hat{\mathbf{k}}_\perp$
components ($\hat{\mathbf{k}}_\perp$ is a unit vector perpendicular to
$\mathbf{k}$).
The evolution equation for the transversal component is 
 $ \frac{\partial}{\partial s}w_{\mathbf{k},\perp}=-\widetilde{\eta}k^2
  w_{\mathbf{k},\perp}$,
that is decoupled from the rest of Fourier components, that verify
\begin{equation}\label{hidroEqs}
\frac{\partial}{\partial s}\left(\begin{array}{c}
\rho_{\mathbf{k}}\\
w_{\mathbf{k},||}\\
\theta_{\mathbf{k}}
\end{array}\right)=\mathbf{L}
\left(\begin{array}{c}
\rho_{\mathbf{k}}\\
w_{\mathbf{k},||}\\
\theta_{\mathbf{k}}
\end{array}\right), 
\end{equation}
with
\begin{equation}\label{hidroMatrix}
\mathbf{L}=-\left(\begin{array}{ccc}
0 & ik & 0 \\
\frac{i}{2}k & \widetilde{\eta}k^2 & \frac{i}{2}k\\
2\beta+\widetilde{\mu}k^2 & ik & \beta+\widetilde{\kappa}k^2 
\end{array}
\right), 
\end{equation}
where the dimensionless transport coefficients, 
$\widetilde{\eta}\equiv\frac{\sigma}{\sqrt{2mT_s}}\eta_s$ ,
$\widetilde{\kappa}\equiv\sqrt{\frac{m}{2T_s}}\sigma\kappa_s$, 
$\widetilde{\mu}\equiv\sqrt{\frac{m}{2}}\frac{n_s\sigma}{T_s^{3/2}}\mu_s$, 
have been introduced (the subindex $s$ in the bare transport coefficients
indicates that they are evaluated in the homogeneous stationary
state). To obtain Eq. (\ref{hidroEqs}), the needed quantities 
$\frac{\partial G (n_s,T_s)}{\partial
  T}$ and $\frac{\partial G (n_s,T_s)}{\partial n}$ 
have been identified taking into account the analysis of the
homogeneous phase made previously. In effect, 
$\frac{\partial G (n_s,T_s)}{\partial
  T }=n_s\sigma\sqrt{\frac{2T_s}{m}}\beta$ due to
Eq. (\ref{deltaTEq}), while $\frac{\partial G (n_s,T_s)}{\partial n}$
have been calculated using that $G(n_s,T_s)=0$, so that 
$\frac{\partial G (n_s,T_s)}{\partial n}=\frac{2T_s}{n_s}
\frac{\partial G (n_s,T_s)}{\partial T}$. 

\begin{figure}
\begin{center}
\includegraphics[angle=0,width=0.7\linewidth,clip]{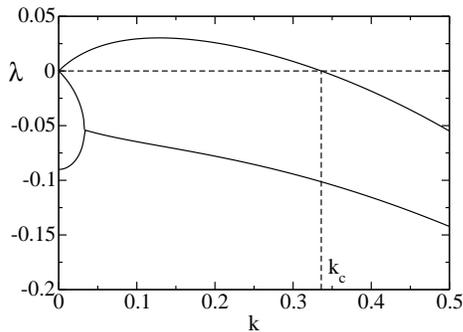}
\end{center}
\caption{Real part of the eigenvalues of $\mathbf{L}$ as functions of
  $k$ for $\epsilon=0.5$ and $\alpha=0.9$. $k$ and $\lambda$ are
  measured in the dimensionless units defined in the main text.  }
\label{dispersion}
\end{figure}

The stability of the system depends on the properties of the matrix
$\mathbf{L}$ given by Eq. (\ref{hidroMatrix}). The eigenvalues, $\lambda$, of
$\mathbf{L}$ are the three roots of the following algebraic equation
\begin{equation}
\lambda^3+[\beta+(\widetilde{\eta}+\widetilde{\kappa})k^2]\lambda^2
+(1+\beta\widetilde{\eta}+\widetilde{\eta}\widetilde{\kappa}k^2)k^2\lambda
-[\beta+(\widetilde{\mu}-\widetilde{\kappa})k^2]\frac{k^2}{2}=0. 
\end{equation}
Here, it is seen that a mode can vanish for a finite
wavenumber, 
$k=k_c\equiv\sqrt{\frac{\beta}{\widetilde{\kappa}-\widetilde{\mu}}}$. In
fact, it can be seen by general arguments that there can be only one
unstable mode for small $k$ \cite{brs13}. This occurs only if the
compressibility in the stationary state is negative,
i.e. $\frac{dp}{dn}(n_s)<0$, being $p(n)\equiv nT_s(n)$, in which case
it goes as
$\lambda\simeq\sqrt{-\frac{1}{2T_s}\frac{dp(n_s)}{dn}}k$ to first
order in $k$. For our
model, the condition for the instability is fulfilled and the unstable
mode goes as $\lambda\simeq\frac{1}{\sqrt{2}}k$. This mode becomes
stable for $k>k_c$ due to heat conduction (note that $k_c$ depends
also on $\widetilde{\mu}$). In
Fig. \ref{dispersion}, the real part of the eigenvalues of $\mathbf{L}$ are plotted for
$\epsilon=0.5$ and $\alpha=0.9$ as functions of $k$. The unstable mode
is clearly identified. For the transport coefficients, we have taken
their 2D elastic values \cite{g71},
consistently with the previous analysis. Hence, a
finite system of area $L\times L$ can be stable or unstable depending
if $L<L_c\equiv\frac{2\pi}{n_s\sigma k_c}$ or $L>L_c$
respectively. 
Note the strong analogy with the instability of the free
cooling case \cite{gz93} although, as it is seen in
Fig. \ref{dispersion}, the structure of the modes is very different. 

The theory developed here can be generalized for moderated 
densities. At the Enskog level, position correlations can be taken into
account in an effective way by multiplying all the terms excepting the
wall contribution in Eqs (\ref{ecT}) and (\ref{ecTz}) by the pair
correlation function at contact, $g_2(n_s)$. In this case, performing a
similar analysis, the expression for the critical wavenumber is 
$k_c=\sqrt{\frac{\beta
    g_2(n_s)}{\widetilde{\kappa}-\widetilde{\mu}}\left(1+2n_s\frac{d\log
      g_2(n_s)}{dn}\right)}$. We have performed MD simulations to
check the validity of this expression. Concretely, we have taken
$\alpha=0.9$, $\epsilon=0.5$ and two different densities,
$n_s\sigma^2=0.15$ and $n_s\sigma^2=0.3$. For the first density and taking
the approximate expression for $g_2$ given in \cite{g71}, the
theoretical prediction is such that the
system is supposed to be unstable above a number of particles 
$N_c\equiv\frac{4\pi^2}{n_s\sigma^2k_c^2}\sim 1650$. We have performed
MD simulations finding that for 
$N=1300$ and $1400$ the 
system is stable, for $N=1700$ and $1800$ unstable, while it is hard
to say anything reliable for $N=1500$
and $1600$ as the system fluctuates from the
homogeneous to the inhomogeneous phase. In the second density, $N_c\sim
620$, and for $N=500$ and $600$ the
steady state was observed to be stable, while for $N=800$ and $900$ the
system developed the instability. For $N=700$ no conclusion could be
reached from the simulations. Hence, we can say that the theory
predicts well the stability/instability of the system.  

To sum up, we have introduced a simple model that allows to explain 
from a microscopic point of view the origin of the instability in Q2D
granular systems. Essentially, the idea is that 
there is one mechanism that destabilizes the system in the steady
state (negative
compressibility) and another one that stabilizes it (dissipation due to
heat conduction). For small gradients, the first mechanism dominates the
second one, making the system unstable, while for large gradients, the
perturbation decays faster and the system is linearly stable. The fact 
that the compressibility is negative can be understood from the
particular way in which energy is injected in the vertical direction and
transferred to the horizontal direction through collisions. Hence, the
instability is more a question of ``size'' that of density and it is
expected that a critical length can be identified in the experiments. 
Of course,
the instability can be tuned by many other aspects such as friction
with the walls, inelasticity of particle-wall collisions, or gravity
to mention a few but, in our opinion, the essential ingredients have
been 
identified. On the other hand, the coexistence between the solid and
gas phases has not been treated as the formalism is no longer valid
for high densities. Nevertheless, it seems that the simplicity of the
model would allow to make progress in this direction. 

\section{acknowledgments}
This research was supported by the Ministerio de Econom\'ia, Industria
y Competitividad (Spain) through Grant No. FIS2017-87117-P (partially financed
by FEDER funds).

\end{document}